\documentclass[12pt]{article}
\usepackage{amsmath,amssymb,amsfonts,amsthm}
\usepackage[english]{babel}
\usepackage{graphicx}
\usepackage{float}
\newcommand{\be}{\begin{equation}}
\newcommand{\ee}{\end{equation}}
\newcommand{\bea}{\begin{eqnarray}}
\newcommand{\eea}{\end{eqnarray}}
\newcommand{\nn}{\nonumber \\}
\newcommand{\p}[1]{(\ref{#1})}
\newcommand{\lb}{\label}

\begin{document}
\begin{titlepage}

\vfill
\vfill
\vspace*{2cm}
\begin{center}
\baselineskip=16pt {\Large\bf Super K\"ahler oscillator from $SU(2|1)$ superspace}

\vskip 0.3cm {\large {\sl }} \vskip 10.mm {\bf $\;$  E. Ivanov$^{\,a}$,  $\;$ S. Sidorov$^{\,b}$
}
\vspace{1cm}

{\it Bogoliubov Laboratory of Theoretical Physics, JINR, \\
141980 Dubna, Moscow Region, Russia\\
}
\end{center}
\vfill

\par
\begin{center}
{\bf Abstract}
\end{center}
We construct a new version of the worldline $SU(2|1)$ superspace as a deformation of the standard
${\cal N}=4, d=1$ superspace and show that it naturally provides off- and on-shell description of general
supersymmetric K\"ahler oscillator model  considered earlier at the classical level within the Hamiltonian approach.
The basic object is a generalized chiral $SU(2|1), d=1$ superfield with the off-shell field content ${\bf (2, 4, 2)}$.
The frequency of the oscillator and the strength of external magnetic field are defined by two parameters:
the contraction parameter $m$ and the new parameter $\lambda$ which reflects the freedom in defining
the chiral $SU(2|1), d=1$ superspace. We treat both classical and quantum cases.
\vspace{2cm}


\begin{quote}
\vfill \vfill \vfill \vfill \vfill \hrule width 5.cm \vskip 2.mm
{\small
\noindent $^a$ eivanov@theor.jinr.ru\\
\noindent $^b$ sidorovstepan88@gmail.com\\
}
\end{quote}
\end{titlepage}

\setcounter{footnote}{0}

\setcounter{page}{1}

\numberwithin{equation}{section}

\section{Introduction}
Recently, we proposed a new type of supersymmetric quantum mechanics (SQM) based on the worldline realization of the supergroup $SU(2|1)$
in the appropriate ${\cal N}=4, d=1$ superspace \cite{DSQM}\footnote{This study was partly motivated by a recent interest in the theories
possessing curved rigid analogs of the Poincar\'e supersymmetry (see, e.g., \cite{FS,DFS} and refs. therein).}.
The corresponding SQM models are deformations of the standard ${\cal N}=4$ SQM models
by the intrinsic mass parameter $m$, such that the ${\cal N}=4$ models are reproduced in the limit $m=0\,$.
The $SU(2|1)$ supersymmetry acts on the worldine multiplets
of the same off-shell dimension as in the standard ${\cal N}=4, d=1$ supersymmetry. In \cite{DSQM} we considered
models associated with the off-shell $d=1$
multiplets $({\bf 1, 4, 3})$ and $({\bf 2, 4, 2})$. In the first type of models the $SU(2|1)$ supersymmetry
is recognized on shell as the
``weak supersymmetry'' of ref. \cite{WS}, while the models of the second type provide some interesting deformations
of the standard ${\cal N}=4, d=1$ K\"ahler
sigma models \cite{242stand}, such that they contain proper couplings to the external magnetic field
and intrinsic potential terms related
to the K\"ahler potential and vanishing in the $m=0$ limit. Another peculiarity of these models
is the very restrictive form of the extra superpotential terms.
No such restrictions appear in the $({\bf 2, 4, 2})$ SQM models with the standard ${\cal N}=4, d=1$ supersymmetry.

On the other hand, there is a different class of mechanical systems with $SU(2|1)$ as the underlying supersymmetry, known
in the literature as supersymmetric K\"ahler oscillator \cite{BN3,BN4}. These models deal with  the same number
of physical bosonic and fermionic variables (multiples of $2 + 4$) as our $({\bf 2, 4, 2})$ $SU(2|1)$ models, but essentially
differ in the structure of the relevant supercharges and Hamiltonian. They involve 2 deformation parameters:
the strength of a constant magnetic field $B$ and the frequency of oscillator $\omega$. While the first parameter can be
identified with the contraction parameter $m$, it was unclear how to interpret the second parameter
within our $SU(2|1), d=1$ superspace approach.

In the present note we explain how to incorporate the models of \cite{BN3,BN4} into the
$SU(2|1), d=1$ superspace formalism. To this end, one needs to define the $SU(2|1), d=1$ superspace in
a way different from that employed in \cite{DSQM}. The basic difference is that the Hamiltonian of the system should
be identified with the whole intrinsic $U(1)$ charge of the $SU(2|1)$ superalgebra, but not with the central charge
as in \cite{DSQM}.

The new version of the $SU(2|1), d=1$ superspace still contains the chiral subspace having twice as less Grassmann coordinates, but the definition of this subspace now reveals a freedom parametrized by a new parameter $\lambda$.
Just this freedom is responsible for the appearance of the correct K\"ahler potential term in the Hamiltonian with the
frequency $\omega \sim \sin 2\lambda\,$. The extra superpotentials also exhibit more freedom compared to those appearing in the
``old'' $({\bf 2, 4, 2})$ $SU(2|1)$ models: they are parametrized by an arbitrary holomorphic function of the complex
bosonic variables $z^a$.

The paper is organized as follows. In Sect.~2, 3 we define the new version of the $SU(2|1)$ superspace and explain its relation to the superspace
 of \cite{DSQM}, as well as the origin of the $\lambda$ freedom mentioned above. In Sect.~4 we introduce the generalized chiral $SU(2|1)$ superfield
 which involves the parameter $\lambda$ appearing in the off-shell transformations of the component fields. The general
 off- and on-shell Lagrangians of these chiral superfields,
including the extra superpotentials, are constructed in Sect.~5, starting from the superfield approach.  We also compute the relevant Hamiltonian,
supercharges and other $SU(2|1)$ generators, in both the classical and the quantum cases, and check that they indeed form the
algebra $su(2|1)$. Concluding remarks are collected in Sect.~6.

\setcounter{equation}{0}
\section{New $SU(2|1)$ superspace}
\subsection{The $su(2|1)$ superalgebra and its $u(1)$ extension}
The standard form of the superalgebra $su(2|1)$ is:
\bea
    &&\lbrace Q^{i}, \bar{Q}_{j}\rbrace = 2m I^i_j +2\delta^i_j \tilde H,\qquad\left[I^i_j,  I^k_l\right]
    = \delta^k_j I^i_l - \delta^i_l I^k_j\,,\nn
    &&\left[I^i_j, \bar{Q}_{l}\right] = \frac{1}{2}\delta^i_j\bar{Q}_{l}-\delta^i_l\bar{Q}_{j}\, ,\qquad \left[I^i_j, Q^{k}\right]
    = \delta^k_j Q^{i} - \frac{1}{2}\delta^i_j Q^{k},\nn
    &&\left[\tilde H, \bar{Q}_{l}\right]=\frac{m}{2}\bar{Q}_{l}\,,\qquad \left[\tilde H, Q^{k}\right]=-\frac{m}{2}Q^{k},    \label{algebra1}
\eea
all other (anti)commutators vanishing. The generators $I^i_j$ generate $SU(2)$ symmetry, while the mass-dimension generator
$\tilde H$ is $U(1)$ symmetry generator.
The arbitrary mass parameter $m$ is treated as a contraction parameter: sending $m\rightarrow 0$
leads to the ${\cal N}=4$, $d=1$ Poincar\'e superalgebra. In the limit $m = 0$, $\tilde H$ become
the canonical Hamiltonian and the generators $I^i_j$ become the outer $SU(2)$ automorphism generators.

To see the difference between the $SU(2|1), d=1$ superspace introduced in \cite{DSQM} and its modification we are going to deal with here, it is
instructive to extend \p{algebra1} by the external $U(1)$ automorphism symmetry generator $F$ which has non-zero commutation relations only
with the supercharges \cite{FS}:
\bea
    \left[F, \bar{Q}_{l}\right]=-\frac{1}{2}\,\bar{Q}_{l}\,,\qquad \left[F, Q^{k}\right]=\frac{1}{2}\,Q^{k}. \lb{Fgen}
\eea
The redefinition $\tilde H \equiv H - m F $ brings the extended superalgebra $su(2|1)\oplus u(1)_{\rm ext}$
to the form in which it looks as a centrally extended superalgebra $\hat{su}(2|1)$:
\bea
    &&\lbrace Q^{i}, \bar{Q}_{j}\rbrace = 2m\left( I^i_j -\delta^i_j F\right)+ 2\delta^i_j H\,,\qquad\left[I^i_j,  I^k_l\right]
    = \delta^k_j I^i_l - \delta^i_l I^k_j\,,\nn
    &&\left[I^i_j, \bar{Q}_{l}\right] = \frac{1}{2}\delta^i_j\bar{Q}_{l}-\delta^i_l\bar{Q}_{j}\, ,\qquad \left[I^i_j, Q^{k}\right]
    = \delta^k_j Q^{i} - \frac{1}{2}\delta^i_j Q^{k},\nn
    &&\left[F, \bar{Q}_{l}\right]=-\frac{1}{2}\,\bar{Q}_{l}\,,\qquad \left[F, Q^{k}\right]=\frac{1}{2}\,Q^{k}.\label{algebra}
\eea
In the new basis $(F, H)$ the generator $F$ becomes the internal $U(1)$ generator, while $H$ commutes with all generators
and so can be treated as a central charge.

\subsection{Two $d=1$ supercosets of $SU(2|1)$}
The  $SU(2|1), d=1$ superspace introduced in \cite{DSQM} can be identified with the following coset
of the extended superalgebra $su(2|1)\oplus u(1)_{\rm ext}$ in the basis ${H, F}$:
\be
    \frac{SU(2|1)\rtimes U(1)_{\rm ext}}{SU(2)\times U(1)_{\rm int}}\, \sim \,
    \frac{\{Q^{i},\bar{Q}_{j}, H, F, I^i_j \}}{\{I^i_j, F\}}\,.\lb{oldSP}
\ee
Its main property is that the time coordinate $t$  is associated as the coset parameter with the central charge generator $H$.
This generator commutes
with all other ones and, in the corresponding SQM models,  can naturally be identified with the Hamiltonian.

In the present paper we will deal with another $SU(2|1)$ coset
\be
    \frac{SU(2|1)\rtimes U(1)_{\rm ext}}{SU(2)\times U(1)_{\rm ext}} = \frac{SU(2|1)}{SU(2)}\, \sim \,
    \frac{\{Q^{i},\bar{Q}_{j},\tilde H,I^i_j \}}{\{I^i_j\}}\,.\lb{newSP}
\ee
The role of the Hamiltonian is now played by the full internal $U(1)$ generator $\tilde H$. Though it does
not commute with the supercharges,
the Noether expressions for the latter are still conserved due to the presence of explicit dependence on $t$ in them.
This situation is reminiscent of what happens, e.g.,  in the conformal and superconformal mechanics \cite{superc}.

\subsection{The second $SU(2)$ group}
For what follows, it will be useful to have a different notation for the superalgebra \p{algebra}.
Introducing supercharges with the doublet $SU(2)_{\rm ext} \times SU(2)$ indices as\footnote{The doublet indices are raised and lowered
in the standard way by the $\varepsilon$ symbols, e.g., $Q_i = \varepsilon_{ik}Q^k, \; \bar{Q}^i = \varepsilon^{ik}\bar Q_k\,, \;\;
\varepsilon_{12}= -\varepsilon^{12} =1\,$.}
\bea
    Q_{1i}\equiv Q_{i}\,,\quad Q_{2i}\equiv \bar{Q}_{i} \,,\qquad\left(Q_{bj}\right)^\dagger = Q^{bj}, \label{charges}
\eea
the relations \p{algebra} can be rewritten in the ``quasi'' $SU(2)_{\rm ext} \times SU(2)$ covariant form as
\bea
    &&\lbrace Q_{ai}, Q_{bj}\rbrace =2 i m c_{ab}\, I_{ij} +2\varepsilon_{ij}\,\varepsilon_{ab}\,\tilde H,\qquad\left[I_{ij},  I_{kl}\right]
    = \varepsilon_{kj}\, I_{il} - \varepsilon_{il}\, I_{kj}\,,\nn
    &&\left[I_{ij}, Q_{al}\right] =  -\frac{1}{2}\left( \varepsilon_{il}\, Q_{aj}+\varepsilon_{jl}\, Q_{ai}\right) , \quad
    \left[\tilde H, Q_{ai}\right]=\frac{im}{2}\,c_{a}^{b}\,Q_{bi}\,.\label{alg}
\eea
Here $c_{ab}$ is a constant real triplet of the $SU(2)_{\rm ext}$ symmetry which breaks the latter down to
$U(1)_{\rm ext} \subset SU(2)_{\rm ext}$:
\bea
c_{ab}=c_{ba}\,, \quad  \overline{\left( c_{ab}\right)}=\varepsilon^{ac}\varepsilon^{bd}c_{cd}\,,\; c^{ab}c_{ab}=2.
\eea
Choosing, e.g.,  the $SU(2)_{\rm ext}$ frame in such a way that $c_{12}=c_{21}=-i$, $c_{11}=c_{22}=0$, we reproduce \p{algebra1},
with the external automorphism generator $F$ as the only remnant of this second $SU(2)$. Another frame corresponds
to the choice $c_{12} = 0\,, c_{11} = c_{22} = 1\,,$
which yields the equivalent form of the $su(2|1)$ superalgebra
$$
\lbrace Q_i', Q_j' \rbrace = \lbrace \bar Q_i', \bar Q_j' \rbrace = 2imI_{ij}\,, \quad  \lbrace Q_i', \bar Q_j' \rbrace
= 2\varepsilon_{ij}\tilde{H}\,.  $$
In this frame, the residual $U(1)_{ext}$ automorphism acts as $O(2)$ rotations of the 2-vector $(Q_i', \bar Q_i')$ (the same regards the action
of the internal $U(1)$ charge generator $\tilde{H}$).

Thus the full $SU(2)_{\rm ext}$ symmetry is not the automorphism group of the $su(2|1)$ superalgebra, only
its some $U(1)_{\rm ext}$ ($\sim O(2)_{\rm ext}$)
subgroup is. On the other hand, in the limit $m\rightarrow 0$, this group becomes one of the $SU(2)$ factors of the
full automorphism group $SU(2)\times SU(2)$ of the ${\cal N }=4$, $d=1$ Poincar\'e supersymmetry.
The generalized $SU(2|1)$ chirality we will consider in Sect. 3 is directly related to the existence of this $SU(2)_{\rm ext}$ freedom.
The new parameter $\lambda$ associated with this chirality can be interpreted as a parameter of some $U(1)\subset SU(2)_{ext}$.
Just because $SU(2)_{\rm ext}$ is not an automorphism of $SU(2|1)$, $\lambda$ is a physical parameter at $m\neq 0$.
It becomes removable and hence unphysical
only in the limit $m=0$, when $SU(2)_{\rm ext}$ turns into the automorphism symmetry of the underlying supersymmetry.
\setcounter{equation}{0}
\section{Superspace technicalities}
\subsection{Transformations}
We parametrize the supercoset \p{newSP} by the coordinates $(t,\theta_{i}, \bar{\theta}^{j})$. The time coordinate is associated
with the generator $\tilde H\,$, which plays, in what follows, the role of Hamiltonian. An element of the supercoset is defined as:
\be
g= \exp {( i\tilde\theta_{i}Q^{i}-i\bar{\tilde{\theta}}^{j}\bar{Q}_{j})}\exp{( i t \tilde H)}, \lb{coset1}
\ee
where\footnote{We use the following conventions:
$\bar{\chi}\cdot \zeta=\bar{\chi}^i \zeta_i\,,\; \bar{\chi}\cdot\bar{\zeta}=\bar{\chi}^i\bar{\zeta}_i \,,\; \chi\cdot \zeta =\chi_i \zeta^i\,,
\;(\chi)^2 = \chi\cdot\chi\,, \;(\bar\chi)^2 = \bar\chi\cdot \bar\chi$.}
\bea
\tilde\theta_i =\left[1-\frac{2m}{3}\,(\bar{\theta}\cdot\theta)\right]\theta_i\,,\qquad \overline{\left(\theta_i\right)}=\bar{\theta}^i.
\eea
One can check that, with the order of factors as in \p{coset1}, the superspace coordinates $(t,\theta_{i}, \bar{\theta}^{j})$
transform in the same
manner as those in \cite{DSQM}:
\bea
    &&\delta\theta_{i}=\epsilon_{i}+2m\,(\bar{\epsilon}\cdot\theta)\, \theta_{i}\,,\qquad\delta\bar{\theta}^{j}
    =\bar{\epsilon}^{i}-2m\,(\epsilon\cdot\bar\theta)\,\bar{\theta}^{i},\nn
        &&\delta t=i[(\epsilon\cdot\bar\theta)+ (\bar{\epsilon}\cdot\theta)]\,,\qquad \overline{(\epsilon_i)}=\bar{\epsilon}{}^{i}.\label{tr}
\eea
Then the invariant measure of integration over the new $SU(2|1)$ superspace is defined by the same formula as in \cite{DSQM}:
\bea
d\mu :=   dt\, d^2\theta\, d^2\bar{\theta}\,[1+ 2m\,(\bar{\theta}\cdot\theta)]\,. \label{inv}
\eea

\subsection{Covariant derivatives}
Applying the same general method of Cartan one-forms as in \cite{DSQM}, it is easy to find the corresponding
covariant derivatives:
\bea
    {\cal D}^i&= &e^{-\frac{imt}{2}}\bigg[\left(1+{m}\,(\bar{\theta}\cdot\theta)
    -\frac{3m^2}{4}\,(\bar{\theta}\cdot\theta)^2\right)\frac{\partial}{\partial\theta_i}
    - {m}\,\bar{\theta}^i\theta_j\frac{\partial}{\partial\theta_j}-i\bar{\theta}^i\frac{\partial}{\partial t}\nn
    &&-\, {m}\,\bar{\theta}^j\tilde{I}^i_j
    -\frac{m^2}{2}\,({\bar\theta})^2\,\theta^j\tilde{I}^i_j\bigg],\nn
    \bar{{\cal D}}_j &= &e^{\frac{imt}{2}}\bigg[-\left(1+ {m}\,(\bar{\theta}\cdot\theta)
    -\frac{3m^2}{4}\,(\bar{\theta}\cdot\theta)^2\right)\frac{\partial}{\partial\bar{\theta}^j}
    + {m}\,\bar{\theta}^k\theta_j\frac{\partial}{\partial\bar{\theta}^k}+i\theta_j\frac{\partial}{\partial t}\nn
    &&+\, {m}\,\theta_k\tilde{I}^k_j -\frac{m^2}{2}\,({\theta})^2\,\bar{\theta}_k\tilde{I}^k_j\bigg],\nn
    {\cal D}_t & = & \partial_t\,.\label{cov}
\eea
The objects within the square brackets coincide with the covariant derivatives on the superspace \p{oldSP}, modulo the absence of the matrix parts
with the internal $U(1)$ charge generator which is now out of the stability subgroup. Non-trivial  $U(1)_{\rm int}$ transformations of the ``old''
covariant derivatives are now compensated by the transformations of the time-dependent factors in \p{cov}, so that the new covariant derivatives
are $U(1)_{\rm int}$-inert. The superalgebra of the covariant derivatives mimics the superalgebra $su(2|1)\,$:
\bea
    &&\lbrace {\cal D}^i, \bar{{\cal D}}_j\rbrace = 2m\tilde{I}^i_j +2i\delta^i_j{\cal D}_t \,,\qquad
    \left[\tilde{I}^i_j, \tilde{I}^k_l\right] = \delta^i_l \tilde{I}^k_j -\delta^k_j \tilde{I}^i_l  \,,\nn
    &&\left[\tilde{I}^i_j, \bar{{\cal D}}_{l}\right] = \delta^i_l\bar{{\cal D}}_{j}-\frac{1}{2}\delta^i_j\bar{{\cal D}}_{l}\, ,\qquad
    \left[\tilde{I}^i_j, {\cal D}^{k}\right] =  \frac{1}{2}\delta^i_j {\cal D}^{k} -\delta^k_j {\cal D}^{i} ,\nn
    &&\left[{\cal D}_t, \bar{{\cal D}}_{l}\right]=\frac{im}{2}\bar{{\cal D}}_{l}\,,\qquad \left[{\cal D}_t, {\cal D}^{k}\right]
    =-\frac{im}{2}{\cal D}^{k}.
\eea

Note that one can ascribe to ${\cal D}^i$ and $\bar{\cal D}_i$ (as well as to $\theta_i$ and $\bar\theta^i$) the opposite charges with respect
to the $U(1)_{\rm ext}$ generator $F$ which can be formally kept in the stability subgroup in \p{newSP}. However, contrary
to the internal generator $F$ in \p{oldSP}, the automorphism generator $F$ in \p{newSP} never appears in the r.h.s.
of the anticommutators of the supercharges. In the first case, the internal $U(1)$ invariance is the necessary consequence
of supersymmetry and therefore should be respected by any corresponding SQM model. In the second case, the automorphism $U(1)$
symmetry is not automatically implied by supersymmetry. So
this $U(1)$ invariance is merely an additional possible restriction on the SQM models: one may impose it, or may not.

\setcounter{equation}{0}
\section{Generalized chiral $SU(2|1)$ superfields}
The standard form of the chiral and antichiral conditions is
\bea
 {\rm (a)}\;\;   \bar{{\cal D}}_i\Phi =0\,,\qquad  {\rm (b)} \;\;{\cal D}^i\bar\Phi =0\,. \label{ch}
\eea
In the framework of the superspace approach of \cite{DSQM}, it was the only option to describe the multiplet $({\bf 2,4,2})$.
It is uniquely specified by the covariance with respect to the stability subgroup $U(2) = SU(2)\times U(1)_{int}\,$. In the case we consider here,
the stability subgroup is actually $SU(2)$ (modulo the unessential automorphism $U(1)_{ext}$ group). Capitalizing on that, we can generalize
the chiral condition \p{ch}  as
\bea
   {\rm (a)}\;\;  \bar{\tilde{{\cal D}}}_i \Phi =0\,,\qquad {\rm (b)} \;\;\tilde{{\cal D}}^i\bar{\Phi}=0\,,\label{ch1}
\eea
where
\bea
    \bar{\tilde{{\cal D}}}_i = \cos{\lambda}\,\bar{{\cal D}}_i -\sin{\lambda}\,{\cal D}_i \,,\qquad \tilde{{\cal D}}^i
    =\cos{\lambda}\,{\cal D}^i +\sin{\lambda}\,\bar{{\cal D}}^i\,.\label{rotated}
\eea
Clearly, in the approach based on the superspace \p{oldSP}, the constraints \p{ch1} are not covariant under $U(1)_{\rm int}$
which multiplies ${\cal D}_i$ and $\bar{\cal D}_i$ by the mutually conjugated phase factors. In the case under consideration
${\cal D}_i$ and $\bar{\cal D}_i$ undergo no any supersymmetry-induced $U(1)$ phase transformations, and so the conditions \p{ch1} are $SU(2|1)$
covariant at any $\lambda$. The linear combinations \p{rotated} can be interpreted as a result of the rotation of the $SU(2)_{\rm ext}$ doublet
${\cal D}_{ai} := ({\cal D}_i, \bar{\cal D}_i)$ by some one-parameter subgroup of $SU(2)_{\rm ext}$ acting on the doublet index $a$ (this $SU(2)$
 group is the same as in Sect. 2.3). Since this subgroup is not an automorphism of \p{algebra}, the $\lambda$ dependence  cannot be removed
 from \p{ch1}, \p{rotated} by any redefinition of the Grassmann coordinates $\theta_i, \bar\theta^k$. This is possible only in the limit $m=0$, when
 $SU(2)_{\rm ext}$  becomes the automorphism group of the ${\cal N}=4, d=1$ superalgebra.

The conditions \p{ch1} amounts to the existence of the left and right chiral subspaces:
\bea
    (t_L,\hat\theta_i),\qquad  (t_R, \bar{\hat{\theta}}^i)\,,
\eea
where
\bea
    && t_L = t +i\cos{2\lambda}\,(\bar{\theta}\cdot\theta) + \frac{i}{2}\sin{2\lambda} \left[(\bar\theta)^2 e^{-i m t}
    + (\theta)^2 e^{i m t}\right] -im\cos{2\lambda}\,(\bar{\theta}\cdot\theta)^2,\nn
    &&\hat{\theta}_i = (\cos{\lambda}\,\theta_i e^{\frac{i}{2}m t}+\sin{\lambda}\,
    \bar{\theta}_l e^{-\frac{i}{2}m t})
    \left[1 -\frac{m}{2}\,(\bar{\theta}\cdot\theta)\right] \label{subs}
\eea
(the right subspace coordinates are obtained via complex conjugation). Indeed, in the basis $(t_L, \hat\theta_i, \bar{\hat\theta}^k)$
the constraint (\ref{ch1}a) is reduced to the form $\partial_{\bar{\hat\theta}^k}{\Phi } = 0 \,\Rightarrow\, {\Phi } = \varphi(t_L,\hat\theta_i)$.
As it should be, the coordinate set \p{subs} is closed under the $SU(2|1)$ transformations
\bea
    &&\delta\hat{\theta}_{i}=\cos{\lambda}\,[\epsilon_{i}e^{\frac{i}{2}m t_L} +
    m(\bar{\epsilon}\cdot\hat{\theta})\hat{\theta}_{i} e^{-\frac{i}{2}m t_L}]
    +\sin{\lambda}\,[\bar{\epsilon}_k e^{-\frac{i}{2}m t_L}
    -m(\epsilon \cdot\hat{\theta})\hat{\theta}_{i} e^{\frac{i}{2}m t_L}]\,,\nn
    && \delta t_L=2i[\cos{\lambda}(\bar{\epsilon}\cdot\hat{\theta}) e^{-\frac{i}{2}m t_L}
    + \sin{\lambda}(\epsilon \cdot\hat{\theta})e^{\frac{i}{2}m t_L}]\,.\label{newtr}
\eea
Eqs. \p{subs}, \p{newtr} suggest that in the $SU(2|1)$ case there exists a family of the non-equivalent chiral subspaces parametrized
by $\lambda$. It is straightforward  to show that in the $m=0$ case the $\lambda$ dependence can be entirely absorbed into
the proper redefinition of the Grassmann coordinates and the supersymmetry transformation parameters. This is not possible at $m\neq 0$,
i.e. in the $SU(2|1)$ case.

The rotated covariant derivatives \p{rotated} satisfy the relations
\bea
    \{\tilde{\bar{{\cal D}}}_k ,\tilde{\bar{{\cal D}}}_j \} = -2 m\sin{2\lambda} \,\tilde{I}_{ij}\, \qquad \mbox{and \;c.c.},
\eea
which means that the chiral superfields $\varphi$ cannot carry any external $SU(2)$ index, i.e. they must be $SU(2)$ singlets. The
left chiral superfield $\varphi (t_L,\hat\theta_i)$ as a solution of eq. \p{ch1} is given by the standard expansion
\bea
    \varphi (t_L,\hat\theta_i)=z+\sqrt{2}\,\hat{\theta}_k\xi^k +(\hat\theta)^2 B ,\qquad \overline{\left(\xi^i\right)}= \bar{\xi}_i\,. \label{leftch}
\eea
The odd transformations \p{newtr} induce the following off-shell transformations of the component fields in \p{leftch}:
\bea
    &&\delta z  =-\sqrt{2}\cos{\lambda}\left(\epsilon\cdot\xi\right) e^{\frac{i}{2}m t}+\sqrt{2}\sin{\lambda}\,
    \left(\bar{\epsilon}\cdot\xi\right) e^{-\frac{i}{2}m t},\nn
    &&\delta \xi^i =  \sqrt{2}\, \bar{\epsilon}^i\,[i\cos{\lambda}\,\dot z-\sin{\lambda} \,B]\,
    e^{-\frac{i}{2}m t}-\sqrt{2}\,\epsilon^i\,[i\sin{\lambda}\,\dot z+ \cos{\lambda} \,B]\, e^{\frac{i}{2}m t},\nn
    &&\delta B = \sqrt{2}\cos{\lambda} \,[i(\bar{\epsilon}\cdot\dot{\xi}\,)+\frac{m}{2}(\bar{\epsilon}\cdot\xi)]\,
    e^{-\frac{i}{2}m t}+\sqrt{2}\sin{\lambda}\,[i(\epsilon\cdot\dot{\xi}\,)-\frac{m}{2}(\epsilon\cdot\xi)]\,e^{\frac{i}{2}m t}.\label{offsh}
\eea
In the special case with $\sin\lambda =0$, the $\theta$ expansion of $\varphi$ takes the form,
\bea
    \varphi (t_L,\hat\theta_i)=z+\sqrt{2}\,\hat{\theta}_k\xi^k +(\hat\theta)^2 B
    =z+\sqrt{2}\,e^{\frac{i}{2}m t_L}\theta_k\xi^k +e^{im t_L}\left(\theta\right)^2 B.
\eea
After the field redefinitions
\bea
    z\rightarrow z,\;\;\; e^{\frac{i}{2}m t_L} \xi^i \;\rightarrow\; \xi^i ,\;\;\;e^{im t_L} B\;\rightarrow \;B\,,
\eea
this particular  $\varphi(t_L, \hat\theta_i)$ is recognized as the chiral superfield of ref. \cite{DSQM}, with zero $U(1)_{\rm int}$ charge.

Finally, note that the one-parameter set of the chirality conditions \p{ch1}, \p{rotated} is in fact the most general linear set.
One could start in (\ref{ch1}a) with the general linear combination of $\bar{\cal D}_i$ and ${\cal D}_i$
involving two complex coefficients. Then, using the freedom of multiplying such a constraint by an arbitrary non-zero factor and making
the proper phase rotation of the Grassmann coordinates, one can reduce it just to the form (\ref{ch1}a).

\setcounter{equation}{0}
\section{Supersymmetric K\"ahler oscillator}
\subsection{The general kinetic Lagrangian}
The most general sigma-model part of the $SU(2|1)$ invariant action of the generalized chiral superfields $\varphi^a(t_L, \hat\theta),
\,a= 1, \ldots N\,,$ is specified
by an arbitrary K\"ahler potential $ f(\varphi^a , \bar{\varphi}^{\bar a})$:
\bea
    S_{\rm kin}= \int dt\,{\cal L}_{\rm kin} = \frac{1}{4}\int d\mu\, f(\varphi^a , \bar{\varphi}^{\bar a})\,.\lb{SuperN}
\eea

In the generic case we will consider only the corresponding bosonic Lagrangian, since working out the fermionic terms is straightforward.
This Lagrangian reads:
\bea
    \left.{\cal L}_{\rm kin}\right|=g_{\bar{a}b}\dot{\bar{z}}^{\bar{a}}\dot{z}^b -\frac{i}{2}m\cos{2\lambda}
    \left(\dot{\bar{z}}^{\bar{a}} f_{\bar{a}} -\dot{z}^a f_{a}\right)
    -\frac{m}{2}\sin{2\lambda}\,[\bar{B}^{\bar{a}} f_{\bar{a}}+f_{a}B^a]+g_{\bar{a}b}\bar{B}^{\bar{a}}B^b\,,
\eea
where $g_{\bar a b} = \partial_{{\bar a}}\partial_{b} f(z, \bar z)\,, \; f_{a} = \partial_{a}f(z, \bar z)\,$.
After elimination of the auxiliary fields $\bar{B}^{\bar{a}}$ and $B^b$, the bosonic Lagrangian becomes
\bea
     \left.{\cal L}_{\rm kin}^{\rm on}\right|=g_{\bar{a}b}\dot{\bar{z}}^{\bar{a}}\dot{z}^b -\frac{i}{2}m\cos{2\lambda}
     (\dot{\bar{z}}^{\bar{a}} f_{\bar{a}} -\dot{z}^a f_{a})- \frac{m^2}{4}g^{\bar{a}b}\sin^2{2\lambda}\,
     f_{\bar{a}}f_b\,,\quad  g^{\bar{a}b} :=(g_{\bar{a}b})^{-1}\,.\label{bos}
\eea
It is recognized as the Lagrangian of the K\"ahler oscillator (the sum of the 1st and the 3d terms) extended   by a coupling to an external magnetic
field (WZ-term). All three terms arise from the single superfield term \p{SuperN}. Both the strength of the magnetic field
and the oscillator frequency
prove to be expressed through the same parameter $\lambda\,$.
One more interesting feature of the Lagrangian \p{bos} is that either WZ-term or K\"ahler oscillator potential $\sim g^{\bar{a}b}f_{\bar{a}}f_b$
can be eliminated
by the proper choice of $\lambda$ ($\cos{2\lambda} =0$ or $\sin{2\lambda} =0\,$), for an arbitrary $f(z,\bar z)$. In the (${\bf 2, 4, 2}$)
Lagrangian of
ref. \cite{DSQM} the WZ-term vanishes only for some very special choices of the K\"ahler potential and non-zero intrinsic
$U(1)$ charge of the chiral
superfields.

Performing Legendre transformation, we find the bosonic Hamiltonian
\bea
     {\cal H}_{\rm bos} =g^{\bar{a }b} \left[\left( p_b -\frac{i }{2}m\cos{2\lambda}\, f_b\right)
     \left(p_{\bar{a}}+\frac{i }{2}m\cos{2\lambda}\,f_{\bar{a}} \right)+ \frac{m^2}{4}\sin^2{2\lambda}\, f_{\bar{a}}f_b\right].\lb{Ham0}
\eea
It is just the Hamiltonian of ref. \cite{BN4}, with the relevant magnetic field $B = m\cos{2\lambda}$ and the frequency
$\omega = (m/2)\,\sin{2\lambda}$. In the limit $m=0$ all the $\lambda$ dependent terms in \p{bos} and \p{Ham0} drop out, in accord with
the discussion in Sect. 4.

For simplicity, from now on we will limit our attention to one complex chiral superfield. The corresponding full Lagrangian
reads
\bea
    {\cal L}_{\rm kin} &=& g\dot{\bar{z}}\dot{z}
    + \frac{i }{2}\, g(\bar{\xi}_i\dot{\xi}^i-\dot{\bar{\xi}}_i\xi^i ) +\frac{i}{2}\left(\bar{\xi}\cdot\xi\right)
    [\dot{\bar z}g_{\bar{z}}-\dot{z}g_z] -\frac{1}{2}\left(\xi\right)^2
    \bar{B}g_z-\frac{1}{2}\left(\bar{\xi}\,\right)^2 B g_{\bar{z}}\nn
    &&+ \,g\bar{B}B +\frac{1}{4}\left(\xi\right)^2 \left(\bar{\xi}\,\right)^2
    g_{z\bar{z}}-\frac{i}{2}m\cos{2\lambda}\left(\dot{\bar{z}}f_{\bar{z}} -\dot{z}f_{z}\right)-\frac{m}{2}\sin{2\lambda}
    [\bar{B}f_{\bar{z}}+f_{z}B]\nn
    &&+\,\frac{m}{4}\sin{2\lambda}\left[\left(\xi\right)^2 f_{zz}+\left(\bar{\xi}\,\right)^2
    f_{\bar{z}\bar{z}}\right]-\frac{1}{2}mg\cos{2\lambda}\left(\bar{\xi}\cdot\xi\right).\label{kinterm}
\eea
Eliminating the auxiliary fields, we obtain the on-shell Lagrangian
\bea
    {\cal L}_{\rm kin}^{\rm on} &=& g\dot{\bar{z}}\dot{z}
    + \frac{i }{2}\,g(\bar{\xi}_i\dot{\xi}^i-\dot{\bar{\xi}}_i\xi^i)
    +\frac{i}{2}\left(\bar{\xi}\cdot\xi\right)[\dot{\bar z}g_{\bar{z}}-\dot{z}g_z]+ \frac{1}{4}\left(\xi\right)^2 \left(\bar{\xi}\,\right)^2 R\nn
    &&-\,\frac{i}{2}m\cos{2\lambda}
    (\dot{\bar{z}}f_{\bar{z}} -\dot{z}f_{z}) -
    \frac{m^2}{4}g^{-1}\sin^2{2\lambda}\,f_z f_{\bar{z}}-\frac{1}{2}mg\cos{2\lambda}\left(\bar{\xi}\cdot\xi\right)\nn
    &&+\,\frac{m}{4}\sin{2\lambda}\left(\bar{\xi}\,\right)^2\left(f_{\bar{z}\bar{z}}
    -\frac{f_{\bar{z}}g_{\bar{z}}}{g}\right)+\frac{m}{4}\sin{2\lambda}\left(\xi\right)^2
    \left(f_{zz} -\frac{g_z f_z}{g}\right),\label{kinterm1}
\eea
where
\bea
    R = g_{z\bar{z}}-\frac{g_z g_{\bar{z}}}{g}\,,
\eea
and the corresponding on-shell supersymmetry transformations
\bea
    \delta z &=&-\sqrt{2}\cos{\lambda}\left(\epsilon\cdot\xi\right) e^{\frac{i}{2}m t}+\sqrt{2}\sin{\lambda}\,
    \left(\bar{\epsilon}\cdot\xi\right) e^{-\frac{i}{2}m t},\nn
    \delta \xi^i &=&\sqrt{2}\, \bar{\epsilon}^i\,[i\cos{\lambda}\,\dot z-\frac12 g^{-1}\sin{\lambda}
    ({m}\,\sin{2\lambda}\, f_{\bar{z}}+\left(\xi\right)^2
    g_z)]\, e^{-\frac{i}{2}m t}\nn
    &&-\,\sqrt{2}\,\epsilon^i [i\sin{\lambda}\,\dot z+ \frac12 g^{-1}
    \cos{\lambda}\,({m}\,\sin{2\lambda}\, f_{\bar{z}}+\left(\xi\right)^2 g_z)]\, e^{\frac{i}{2}m t}.
\eea
\subsection{Superpotential}
The chiral subspace integration measure $d\mu_L$ is invariant under the transformations \p{newtr}:
\bea
    d\mu_L =dt_L\, d^2\hat\theta ,\qquad \delta\,d\mu_L  = (\partial_{t_L}\delta t_L
    - \partial_{\hat \theta_i}\delta \hat\theta_i)d\mu_L = 0\,. \label{L}
\eea
Due to  this property, it is possible to define the general external superpotential
\bea
{S}_{\rm pot}= \int dt\, {\cal L}_{\rm pot} = \frac{\tilde m}{4}\left[\int d\mu_L\, U (\varphi)
    +\int d \bar\mu_R\,\bar{U}(\bar\varphi)\right], \lb{Spot}
\eea
where $U(\varphi)$ is an arbitrary holomorphic function. In components, \p{Spot} yields
\bea
    {\cal L}_{\rm pot}=\frac{\tilde m}{4}\,[2\bar{B}\bar{U}_{\bar{z}}(\bar{z})+ 2B U_z(z) - (\xi)^2\, U_{zz}(z)
    -(\bar{\xi})^2\,\bar{U}_{\bar{z}\bar{z}}(\bar{z})].\label{potterm}
\eea
In the limit $m=0$, this expression goes over to the potential term of the standard ${\cal N}=4, d=1$  multiplet $({\bf 2,4,2})$.
The total off-shell Lagrangian ${\cal L}_{\rm kin}+{\cal L}_{\rm pot}$ thus includes the following bosonic potential term
\bea
    -\frac{1}{2}B\left(m\sin{2\lambda}\,f_{z}-\tilde{m}U_z\right)- \frac{1}{2}\bar{B}
    \left(m\sin{2\lambda}\,f_{\bar{z}}-\tilde{m}\bar{U}_{\bar{z}}\right)+ g\bar{B}B.
\eea
The intrinsic bosonic potential in the on-shell Lagrangian \p{kinterm1} is then modified as
\bea
-\frac{1}{4}g^{-1} m^2 \sin^2{2\lambda}\,f_z f_{\bar z}\; \rightarrow\;
-\frac{1}{4}g^{-1}(m\sin{2\lambda}f_z -\tilde{m}\,U_z)(m\sin{2\lambda}f_{\bar z} -\tilde{m}\,\bar{U}_{\bar z})\,.
\eea
The similar modification comes about in terms $\sim \left(\xi\right)^2$ and
$\left(\bar\xi\,\right)^2$. For simplicity, in what follows we will not consider the superpotential contributions.

In the approach based on the superspace \p{oldSP} the chiral integration measure is multiplied by some induced  $U(1)_{\rm int}$
phase factor under the $SU(2|1)$ supersymmetry. This non-invariance of the measure imposes severe restrictions on the form of the admissible
superpotentials \cite{DSQM}. No such restrictions arise in the approach we deal with here.

\subsection{Hamiltonian formalism}
The canonical Hamiltonian corresponding to the Lagrangian \p{kinterm1} reads
\bea
    \tilde H &=&g^{-1}\left( p_z -\frac{i }{2}m\cos{2\lambda}\,
    f_z +\frac{i}{2}g_{z}\,\xi^k \bar{\xi}_k \right)\left(p_{\bar{z}}
    +\frac{i }{2}m\cos{2\lambda}\,f_{\bar{z}}-\frac{i}{2}g_{\bar{z}}\,\xi^k \bar{\xi}_k \right)\nn
    &&
    -\,\frac{1}{4}\left(\xi\right)^2 \left(\bar{\xi}\,\right)^2 R +
    \frac{m^2}{4}g^{-1}\sin^2{2\lambda}\,f_z f_{\bar{z}} +\frac{m}{2}g\cos{2\lambda}\,\xi^k \bar{\xi}_k\nn
    &&-\,\frac{m}{4}\sin{2\lambda}\left(\bar{\xi}\,\right)^2
    \left(f_{\bar{z}\bar{z}}-\frac{f_{\bar{z}}g_{\bar{z}}}{g}\right)-\frac{m}{4}\sin{2\lambda}\,
    \left(\xi\right)^2 \left(f_{zz} -\frac{g_z f_z}{g}\right).
    \label{canH}
\eea
Taking into account that we deal with one complex chiral superfield, it can be checked that this Hamiltonian
coincides with ${\cal H}_0^{SUSY}$ ($N=1$) of ref. \cite{BN4}. The notations used there are related to ours as
\bea
    m\sin{2\lambda} = 2\omega ,\qquad m\cos{2\lambda} = B, \qquad m = 2\Lambda .
\eea
In the particular case $\omega = 0$ ($\lambda = 0$), the Hamiltonian $\tilde H$ reproduces the expression
for the difference of generators $H-mF$ defined in \cite{DSQM}.

The remaining conserved Noether charges are found to be
\bea
      Q^i &=&\sqrt{2}\, e^{\frac{i}{2}mt}\bigg[\cos{\lambda}\,\xi^i \left( p_z -\frac{i }{2}m f_z +\frac{i}{2}g_z\,\xi^k\bar{\xi}_k\right)
      \nn
      &&-\sin{\lambda}\,\bar{\xi}^i\left(p_{\bar{z}}-\frac{i }{2}m f_{\bar{z}} -\frac{i}{2}g_{\bar{z}}\, \xi^k\bar{\xi}_k\right)\bigg] ,\nn
      \bar{Q}_{j} &=& \sqrt{2}\, e^{-\frac{i}{2}mt}\bigg[\cos{\lambda}\,\bar{\xi}_j\left(p_{\bar{z}}+\frac{i }{2}m f_{\bar{z}}
      -\frac{i}{2}g_{\bar{z}}\, \xi^k\bar{\xi}_k\right) \nn
      &&+\sin{\lambda}\,\xi_j \left( p_z +\frac{i }{2}m f_z +\frac{i}{2}g_z\,\xi^k\bar{\xi}_k\right)\bigg],\nn
      I^i_j &=&g\left(\xi^i\bar{\xi}_j  -\frac12 {\delta_j^i}\,\xi^k \bar{\xi}_k \right).\label{Ncharges}
\eea
The Poisson (Dirac) brackets are imposed as
\bea
    \{z, p_z\} =1, \qquad \{\xi^i ,\bar{\xi}_j \} = -i\delta^i_j\, g^{-1},\qquad \{p_z ,\xi^i\} = \frac{1}{2}g_z\, g^{-1}\xi^i .
\eea
We see that the supercharges explicitly depend on time, and this dependence is such that they satisfy the generalized conservation laws:
\bea
    \frac{d}{dt}Q^i=\partial_t Q^i+\{Q^i,\tilde{H}\}=0 ,\qquad \frac{d}{dt}\bar{Q}_{j}
    =\partial_t \bar{Q}_{j}+\{\bar{Q}_{j},\tilde{H}\}=0\,.\label{law}
\eea
This can be easily checked, using the Poisson brackets
\bea
    \{\tilde{H},Q^i\}=\frac{im}{2} Q^i,\qquad \{\tilde{H},\bar{Q}_{j}\}=-\frac{im}{2}\bar{Q}_{j}\,.
\eea

To prepare the system for quantization, it is useful to make the substitution
\be
    (z, \xi^i)\longrightarrow (z, \eta^i),\qquad\eta^i := g^{{1\over 2}}\xi^i. \label{eta-xi}
\ee
In the new variables, the brackets become
\bea
    \{z, p_z\} =1, \qquad \{\eta^i ,\bar{\eta}_j \} = -i\delta^i_j \,,\qquad \{p_z , \eta^i\}
    = \{p_z , \bar{\eta}_j \}=0\,.\lb{brack1}
\eea
The Noether charges \p{Ncharges} and the Hamiltonian \p{canH} are then rewritten as
\bea
    Q^i &=&\sqrt{2}\,g^{-{1\over 2}} \,e^{\frac{i}{2}mt}\,[\cos{\lambda}\,\eta^i \pi(m)
    -\sin{\lambda}\,\bar{\eta}^i\bar{\pi}(-m)]\, ,\nn
    \bar{Q}_{j} &=& \sqrt{2}\, g^{-{1\over 2}} \,
    e^{-\frac{i}{2}mt}\,[\cos{\lambda}\,\bar{\eta}_j\bar{\pi}(m)+\sin{\lambda}\,\eta_j \pi(-m)]\,,\nn
    I^i_j &=&\eta^i\bar{\eta}_j  -\frac12 {\delta_j^i}\,\eta^k \bar{\eta}_k \,,\nn
    \tilde H &=&g^{-1}\Big[ \bar{\pi}\left(m\cos{2\lambda}\right)\pi\left(m\cos{2\lambda}\right)
    - \frac{1}{4} g^{-1} \left({\eta}\right)^2\left(\bar{\eta}\right)^2 R+
    \frac{m^2}{4}\sin^2{2\lambda}\,f_z f_{\bar{z}} \nn
    && -\frac{m}{4}\sin{2\lambda}\left(\bar{\eta}\right)^2  \left(f_{\bar{z}\bar{z}}
    -\frac{f_{\bar{z}}g_{\bar{z}}}{g}\right)-\frac{m}{4}\sin{2\lambda}\left({\eta}\right)^2
    \left({f_{zz}} -\frac{g_z f_z}{g}\right)\nn
    &&-\,\frac{m}{2} g\cos{2\lambda}\,\bar{\eta}_i\eta^i\Big],
\eea
where we defined
\bea
      \pi(m) = p_z -\frac{i }{2}m f_z +\frac{i}{2}g^{-1}g_z\,\eta^k\bar{\eta}_k\,,\qquad
      \bar{\pi}(m)=p_{\bar{z}}+\frac{i }{2}m f_{\bar{z}} -\frac{i}{2}g^{-1}g_{\bar{z}}\, \eta^k\bar{\eta}_k\,.
\eea
\subsection{Quantization}
The brackets \p{brack1} are quantized in the standard way,
\bea
[ \hat z, \hat{p}_z] =i\,, \quad \{\hat{\eta}^i , \hat{\bar{\eta}}_j \}
    = \delta^i_j\,,\quad [\hat{p}_z , \hat{\eta}^i ]
    = [\hat{p}_z , \hat{\bar{\eta}}_j ]=0\,, \qquad \hat p_z = -i\partial_z\,, \;\; \hat{\bar{\eta}}_j
    = \frac{\partial}{\partial \hat{\eta}^j}\,.
\eea
It will be helpful to use the relation
\bea
    \left[\pi(m) ,\bar{\pi}(m)\right] = m g -\frac{1}{2}\,g^{-1} R\,[\eta^k ,\bar{\eta}_k]\, ,
    \eea
where
\bea
\pi(m) = -i\partial_z  -\frac{i }{2}m f_z +\frac{i}{4}g^{-1}g_z\,[\eta^k ,\bar{\eta}_k]\,,\;\;
\bar{\pi}(m)=-i\partial_{\bar{z}}+\frac{i }{2}m f_{\bar{z}} -\frac{i}{4}g^{-1}g_{\bar{z}}\,[\eta^k ,\bar{\eta}_k]\,.
\eea
The general quantization scheme for SQM models was developed in \cite{How}. Following
this procedure, we obtain the quantum operators\footnote{We pass to the picture in which
the supercharges bear no explicit $t$ dependence:
$$
    \psi \rightarrow U \psi,\qquad \left(\hat{Q}^i_{(cov)}\,,\;\hat{\bar{Q}}_{(cov)j}\right)
    \rightarrow U \left(\hat{Q}^i_{(cov)}\,,\; \hat{\bar{Q}}_{(cov)j}\right) U^{-1},\qquad U= e^{it\hat{\tilde{H}}}.
$$}
\bea
    \hat{Q}^i_{(cov)} &=&\sqrt{2}\,g^{-{1\over 2}} [\cos{\lambda}\,\eta^i \pi(m)-\sin{\lambda}\,\bar{\eta}^i\bar{\pi}(-m) ]\, ,\nn
    \hat{\bar{Q}}_{(cov)j} &=& \sqrt{2}\, g^{-{1\over 2}}\,[\cos{\lambda}\,\bar{\eta}_j\bar{\pi}(m)+\sin{\lambda}\,\eta_j \pi(-m)]\,,\nn
    \hat{I}^i_j &=&\eta^i\bar{\eta}_j  -\frac12 {\delta_j^i}\,\eta^k \bar{\eta}_k \,.
\eea
They satisfy the $su(2|1)$ superalgebra \p{algebra1} with the quantum Hamiltonian
\bea
    \hat{\tilde H} &=& g^{-1}\Big[\bar{\pi}\left(m\cos{2\lambda}\right)\pi\left(m\cos{2\lambda}\right)
    - \frac{1}{4} g^{-1} \left({\eta}\right)^2\left(\bar{\eta}\right)^2 R
    +\frac{m^2}{4} \sin^2{2\lambda}\,f_z f_{\bar{z}}   \nn
    &&-\,\frac{m}{4}\sin{2\lambda}\left(\bar{\eta}\right)^2
    \left(f_{\bar{z}\bar{z}}-\frac{f_{\bar{z}}g_{\bar{z}}}{g}\right)-\frac{m}{4}\sin{2\lambda}
    \left({\eta}\right)^2 \left({f_{zz}} -\frac{g_z f_z}{g}\right)\nn
    && +\,\frac{m}{2} g \cos{2\lambda}\,\eta^i\bar{\eta}_i\Big].
\eea

\setcounter{equation}{0}
\section{Concluding remarks}
In this paper, which is a natural continuation of our previous work \cite{DSQM}, we constructed and studied
a new class of the $SU(2|1)$ SQM models
as a deformation of the standard ${\cal N} =4$, $d=1$ SQM based on the chiral supermultiplet $({\bf 2, 4, 2})$.
We found that, as distinct from the standard ${\cal N} =4$ case, in the $SU(2|1)$ case two different definitions of the worldline
superspace are possible. In one of them, the time coset coordinate is associated with the central charge generator, while
in the other the corresponding coset
generator is the full internal $U(1)$ generator. The first option was used in \cite{DSQM}, and the second one was elaborated on
in the present paper. In accordance with the existence of two different $SU(2|1)$ superspaces, there exist two different versions
of the chiral $SU(2|1)$ multiplet $({\bf 2, 4, 2})$ which both are reduced to the same standard linear chiral ${\cal N}=4, d=1$
multiplet in the contraction
limit $m=0$.
The general $SU(2|1)$ SQM models based on the first version were considered in \cite{DSQM}, while here we presented the characteristic
features of the alternative class of the  $SU(2|1)$ SQM models which are built on the second version of the multiplet $({\bf 2, 4, 2})$.
One of these peculiarities is the presence of new dimensionless parameter $\lambda$ which reflects the non-uniqueness of embedding
of the relevant chiral
superspace into the full $SU(2|1)$ superspace.

The SQM models we have deduced from our $SU(2|1)$ superspace approach were earlier constructed in the Hamiltonian on-shell
component formalism in \cite{BN3,BN4} and were called there ``supersymmetric K\"ahler oscillator''. New features provided
by the superfield approach are the off-shell formulation of this class of models: the relevant component Lagrangians with
auxiliary fields (including the general external superpotential) and the new off-shell  $\lambda$-dependent worldline
realization of $SU(2|1)$ supersymmetry. We also presented the full set of the quantum $SU(2|1)$ generators in these models.
All generators can now be systematically derived from the invariant Lagrangians by the standard Noether procedure.

It would be interesting to explicitly solve the corresponding quantum mechanical problems (spectrum of $\tilde{H}$, etc)
for a few simple choices of the underlying K\"ahler potential $f$. Also, it is of interest to look whether the ${\cal N}=4$
SQM models based on the nonlinear version of the multiplet $({\bf 2, 4, 2})$ \cite{nonlin} admit a generalization
to the $SU(2|1)$ case, and which of the two different $SU(2|1)$ superspace approaches is more preferable for such a generalization.
The study of other $SU(2|1)$ counterparts of the basic ${\cal N}=4$ off-shell multiplets and the associated SQM models
by exploiting both types of the  $SU(2|1)$ superspaces is also one of the problems for the future analysis. Note that the $({\bf 1, 4, 3})$
models considered in \cite{DSQM} have the same description within the new $SU(2|1)$ superspace.

\section*{Acknowledgements}
We thank Armen Nersessian for interest in the work and useful discussions. A partial support from the RFBR grants Nr.12-02-00517,
Nr.13-02-91330, the grant DFG LE 838/12-1 and a grant of the Heisenberg-Landau program is kindly acknowledged.

\end{document}